\documentclass[12pt]{article}
\usepackage{epsfig}
\usepackage{color}
\usepackage{amssymb,amsmath}
\usepackage{graphicx}

\newcommand{\bea}{\begin{eqnarray}}
\newcommand{\eea}{\end{eqnarray}}

 \makeatletter
\def\alt{\mathrel{\mathpalette\gl@align<}}
\def\agt{\mathrel{\mathpalette\gl@align>}}
\def\gl@align#1#2{\lower.6ex\vbox{\baselineskip\z@skip\lineskip\z@
\ialign{$\m@th#1\hfil##\hfil$\crcr#2\crcr\sim\crcr}}} \makeatother

\begin{document}
\begin{titlepage}
\hfill
\vbox{
    \halign{#\hfil        \cr
		 UT-HET 069\cr
         KU-PH-011 \cr
		 UT-12-11
         \cr}}
\vspace*{10mm}

\begin{flushright}
\end{flushright}
\vspace*{1.0cm}

\begin{center}
{\Large\bf 
	A light Higgs scenario based on the TeV-scale supersymmetric strong dynamics\\
}
\vspace*{1cm}
{Shinya~Kanemura ${}^{(a)}$\footnote{
kanemu@sci.u-toyama.ac.jp}
},
{\ Tetsuo~Shindou ${}^{(b)}$\footnote{
shindou@cc.kogakuin.ac.jp}
},
{\ Toshifumi~Yamada ${}^{(c)}$\footnote{
toshifumi@hep-th.phys.s.u-tokyo.ac.jp}
}

\vspace*{.2cm}

{${}^{(a)}${\it Department of Physics, University of Toyama\\
	3190 Gofuku, Toyama 930-8555, Japan}
\vskip .15cm
${}^{(b)}${\it Division of Liberal Arts, Kogakuin University\\
1-24-2 Nishi-Shinjuku, Shinjuku-ku, Tokyo 163-8677, Japan}
\vskip .15cm
${}^{(c)}${\it Department of Physics, University of Tokyo\\
7-3-1 Hongo, Bunkyo-ku, Tokyo 113-0033, Japan}
}
\end{center}
\begin{abstract}
	We consider a model based on the supersymmetric QCD theory with
$N_c=2$ and $N_f=3$.
The theory is strongly coupled at the infrared scale $\Lambda_H$.
Its low energy effective theory below $\Lambda_H$
 is described by the supersymmetric standard model
 with the Higgs sector that contains four iso-spin doublets, two neutral iso-spin singlets and
two charged iso-spin singlets.
If $\Lambda_H$ is at the multi-TeV to 10 TeV,
 coupling constants for the F-terms of these composite
 fields are relatively large at the electroweak scale.
Nevertheless, the SM-like Higgs boson is predicted to be as light as 125 GeV
 because these F-terms contribute to
 the mass of the SM-like Higgs boson not at the tree level
 but at the one-loop level.
A large non-decoupling effect due to these F-terms
 appears in the one-loop correction to the triple Higgs boson coupling,
 which amounts to a few tens percent.
Such a non-decoupling property in the Higgs potential
realizes the strong first order phase transition, which is
required for a successful scenario of electroweak baryogenesis.
\end{abstract}
\vskip 0.5cm

\end{titlepage}

\setcounter{footnote}{0}
\baselineskip 18pt
%
\section{Introduction}

\ \ \ Recently, the ATLAS and CMS experiments at the LHC \cite{lhc} have reported
an excess in the gamma-gamma mode at about 125\;GeV, which may be a signal of the
Higgs boson.
In the Standard Model (SM), a light Higgs boson is the evidence of the weakly coupled Higgs sector.
In models for physics beyond the SM, however, the light Higgs boson does not always correspond to
a weakly coupled theory.
The scenario based on little Higgs models \cite{little higgs} is an example of a strongly coupled theory with a light Higgs boson,
 where the Higgs boson arises as a pseudo Nambu-Goldstone boson originating from
 the breaking of some strongly interacting global symmetry at the TeV scale, and
 the Higgs boson mass is kept to be light.

Supersymmetry (SUSY) is one of the most attractive candidates for the physics beyond the SM.
SUSY can solve the gauge hierarchy problem, as 
the quadratic divergence in the radiative correction to the Higgs boson mass is cancelled owing 
to the non-renormalization theorem.
In addition, elementary scalar fields are automatically introduced in the SUSY theory.
The Higgs sector of the minimal SUSY extension of the SM (MSSM) necessarily contains two Higgs doublets.
In the MSSM, the coupling constants in the Higgs potential are determined by the electroweak gauge couplings, 
 and the mass of the SM-like Higgs boson is less than the $Z$ boson mass at the tree level.
With significant radiative corrections due to the large top Yukawa coupling \cite{mssm higgs}, 
the Higgs mass can be pushed up to around 125\;GeV in the case of very large stop masses or
very large left-right stop mixing.

Even within the framework based on SUSY, models with strongly coupled light Higgs boson can be constructed. 
A possible way is to introduce additional R-parity-even chiral superfields which strongly couple to the Higgs sector
 but the F-terms of which do not contribute to the Higgs boson four-point coupling.
In this case, the SM-like Higgs boson is kept to be light.
The strong couplings have rich phenomenological implications.
First, radiative corrections involving the strongly coupled new fields 
 can raise the SM-like Higgs boson mass to 125\;GeV 
 with rather natural choice of the stop masses and mixing.
Second, the strongly coupled fields significantly contribute to the triple SM-like Higgs boson coupling through 
 loop effects, so that it deviates by a few tens percent from the SM prediction 
 \cite{hhh1, hhh2}.
A similar non-decoupling effect tends to enhance the first order electroweak 
 phase transition \cite{hhh-ewbg, 4hdo}.

In the MSSM, it is not easy to make the first order electroweak phase transition (EWPT)
 strong enough to satisfy the sphaleron decoupling condition \cite{sph dec}, which is required 
 by the successful electroweak baryogenesis \cite{ewbg, ewbgmssm}.
The non-decoupling quantum effect of the additional scalar bosons through strong F-term coupling with the Higgs boson
 enhances the first order EWPT, and the difficulty in the MSSM can be significantly relaxed.
The enhancement requires a light SM-like Higgs boson 
 because its mass works as a suppression factor,
 and the 125\;GeV Higgs boson is consistent with the scenario.
For example, the first order EWPT can be enough strong in a SUSY model 
 whose Higgs sector contains four doublets and two charged singlets \cite{4hdo}.
In this model, the coupling among the Higgs boson and the extra bosons in the scalar potential 
 can be taken to be strong, while the quartic self-coupling constants of the Higgs boson are determined only by the D-term;
 i.e., by the electroweak gauge couplings, and the Higgs mass remains light.

When we explore a more fundamental picture of models with strong couplings and a light Higgs boson, 
a quite different landscape from the grand unified theory (GUT) over the grand desert presents 
itself to us.
In models with strong couplings, coupling constants
tend to blow up quickly through the renormalization group running.
With sufficiently strong coupling consistent with successful electroweak baryogenesis, 
the Landau pole appears at the energy scale of multi-TeV to 10 TeV, which is  
much lower than the GUT scale, $O(10^{16})$\;GeV.
There should be a cutoff scale below the energy scale where one comes across 
the Landau pole.
The physics above the cutoff scale might be controlled by some strong dynamics.

The minimal SUSY fat Higgs model \cite{fat} 
is an example of the theory above the cutoff scale of 
a SUSY model with strongly coupled Higgs sector.
This model is based on the strong $\text{SU(2)}_H$ SUSY gauge theory with 
three pairs of doublets, $(T_1,T_2)$, $(T_3,T_4)$ and $(T_5,T_6)$. 
Below the cutoff scale, Higgs iso-spin doublets, $H_u$ and $H_d$, 
and a neutral singlet, $N$, appear as composite fields of $T_i$'s, and 
the other composite fields are decoupled due to their heavy masses.
The low-energy effective theory is described 
by the superpotential $W=\lambda N(H_dH_u-v_0^2)$, where $\lambda$ is a coupling 
constant and $v_0$ is a dimensionful parameter. The effective theory is identical to
the nearly MSSM (nMSSM) \cite{nMSSM}.
The quartic coupling of the Higgs boson gets a $|\lambda|^2$ contribution
through the F-term. Since the Higgs mass is dominated by the 
$|\lambda|^2 v^2$ term with $v\simeq 174$\;GeV, 
the strong coupling tends to enhance the Higgs mass. 
For $\lambda\simeq 2$ and $\tan\beta \simeq 2$, 
the SM-like Higgs boson mass is as large as 200\;GeV, which weakens the first order EWPT
too much.
Extensions of the minimal SUSY fat Higgs model to $N_c=3, N_f=4$ and $N_c=4, N_f=5$
 are discussed in refs.~\cite{beyond fat1}.
Compositeness in SUSY models is discussed in refs.~\cite{beyond fat2}.

In this paper, we propose a new UV-complete model whose low-energy effective theory
 accomodates strong couplings and a light Higgs boson.
The model is based on $\text{SU}(2)_H$ SUSY gauge theory 
 with three pairs of $\text{SU}(2)_H$ doublets.
This model leads to two pairs of $Y=+1/2$ and $Y=-1/2$ 
 iso-spin doublet composite superfields
 as well as several iso-spin singlet composite superfields 
 in the low-energy description,
 which cause flavor changing neutral currents (FCNCs).
To avoid such dangerous FCNCs, we here impose an additional $Z_2$ parity on the model,
 which is unbroken spontaneously.
This $Z_2$ parity can supply a new candidate for dark matter, 
 in addition to the $R$-parity.
In our model, unlike the minimal SUSY fat Higgs model, 
 the $Z_2$-even singlet field $N$ can be heavy enough to
 decouple from the low-energy effective theory, but many composite fields remain light.
The Higgs sector contains two $Z_2$-even doublets which are 
identical to the MSSM-like Higgs doublets and various extra $Z_2$-odd 
superfields such as a pair of doublets, two charged singlets 
and two neutral singlets.
The SM-like Higgs boson mass of 125 GeV can be realized in a natural way
 because the F-terms of the $Z_2$-odd superfields contribute to
 the mass not at the tree level
 but at the one-loop level.
On the other hand, non-decoupling contributions of these fields in radiative corrections
 affect the triple Higgs boson coupling significantly~\cite{hhh2}, and 
 can make the first order EWPT strong enough through the large F-term coupling constants~\cite{4hdo}.

In Section 2,
 we present the basic framework of $SU(2)_H$ SUSY QCD theory 
 with the $Z_2$ parity, whose low-energy description
 gives a composite SUSY Higgs model.
In Section 3,
 we investigate general features of the composite SUSY Higgs sector.
Generally, F-terms involving a large coupling $\lambda$ 
 contribute to the SM-like Higgs boson mass 
 at the tree level, giving rise to a SM-like Higgs boson much heavier than 125 GeV,
 as in the minimal SUSY fat Higgs model.
In Section 4, 
 we consider an extended model where we obtain
 the SM-like Higgs boson as light as 125 GeV in a natural way.
It turns out that its low-energy effective theory describes the phenomenological model 
 in ref.~\cite{4hdo}.
Section 5 is devoted to the conclusion.
\\
\\

\section{Basic Framework}

\ \ \ We introduce a new $SU(2)$ gauge group, denoted by $SU(2)_{H}$,
 and six chiral superfields, denoted by $T_{i} \ (i=1,2,...,6)$, which are doublets of $SU(2)_H$.
$T_{i}$'s are also charged under SM gauge groups $SU(2)_{L} \times U(1)_{Y}$.
We further assign a $Z_{2}$ parity to them.
SM charge and $Z_2$ parity assignments to $T_i$'s are described in Table 1.
\begin{table}
\begin{center}
\begin{tabular}{|c|c|c|c|} \hline
Field                 & $SU(2)_{L}$ & $U(1)_{Y}$ & $Z_{2}$ \\ \hline
$\left(
\begin{array}{c}
T_{1}  \\
T_{2}
\end{array}
\right)$              & 2           & 0          & +       \\ \hline
$T_{3}$               & 1           & +1/2       & +       \\ \hline
$T_{4}$               & 1           & $-$1/2       & +       \\ \hline
$T_{5}$               & 1           & +1/2       & $-$       \\ \hline
$T_{6}$               & 1           & $-$1/2       & $-$       \\ \hline
\end{tabular}
\end{center}
\caption{
SM charge and $Z_2$ parity assignments on the $SU(2)_H$ doublets, $T_i$.
}
\end{table}
Regarding $SU(2)_{H}$ gauge group,
 this model is nothing but the SUSY QCD theory with 2 colors and 3 flavors,
 which is investigated in ref. \cite{Int-Sei}.
$SU(2)_{H}$ gauge coupling becomes strong at an infrared scale, denoted by $\Lambda_{H}$.
The most general tree-level superpotential that is invariant under
 $SU(2)_H \times SU(2)_L \times U(1)_Y \times Z_2$ symmetry
 is given by
\begin{eqnarray}
W_{tree} &=& \frac{1}{2} m_{1} \ {\rm tr} \left[
\left(
\begin{array}{cc}
T^{2}_{2}  & -T^{1}_{2}  \\
-T^{2}_{1} & T^{1}_{1}
\end{array}
\right)
\left(
\begin{array}{cc}
T^{1}_{1} & T^{1}_{2}  \\
T^{2}_{1} & T^{2}_{2}
\end{array}
\right)
\right]
\ + \ m_{3} T_{3} T_{4} \ + \ m_{5} T_{5} T_{6} \nonumber
\\
&=& m_{1} T_{1} T_{2} \ + \ m_{3} T_{3} T_{4} \ + \ m_{5} T_{5} T_{6} \ ,
\end{eqnarray}
 where we assume $m_{1}, m_{3}, m_{5} < \Lambda_{H}$
 so that the theory remains the SUSY QCD theory with $N_c=2$ and $N_f=3$ at the scale $\Lambda_H$.
$T^{a}_{1}$ and $T^{a}_{2} \ (a=1,2)$ respectively
 indicate the $a$-th components of the $SU(2)_{H}$ doublets $T_{1}$ and $T_{2}$.
In the right hand side of the first line,
 the first matrix transforms as $(2^{*}, 2^{*})$ and the second one does as $(2, 2)$
 under $SU(2)_{H} \times SU(2)_{L}$. 
The trace of their product is thus invaraint under $SU(2)_{H} \times SU(2)_{L}$.

Below $\Lambda_{H}$, the theory is described in terms of 
 composite chiral superfields, $M_{ij}^{\prime} = T_{i}T_{j} \ (i \neq j)$, which are singlets of $SU(2)_H$.
Following the arguments in ref. \cite{Int-Sei}, 
 we see that we have the following dynamically generated superpotential below $\Lambda_{H}$:
\begin{eqnarray}
W_{dyn} &=& -\frac{1}{\Lambda^3} \ \epsilon^{ijklmn} \ M_{ij}^{\prime}  M_{kl}^{\prime}  M_{mn}^{\prime} \ ,
\end{eqnarray}
 where $\Lambda$ is some dynamically generated scale.
Thanks to holomorphy, the net effective superpotential is 
 simply the sum of $W_{dyn}$ and $W_{tree}$:
\begin{eqnarray}
W_{eff} &=& W_{dyn} \ + \ W_{tree} \nonumber
\\
&=& W_{dyn} \ + \ m_{1} M_{12}^{\prime}  \ + \ m_{3} M_{34}^{\prime}  \ + \ m_{5} M_{56}^{\prime}  \ .
\end{eqnarray}

Since we cannot determine the K\"{a}hler potential only from holomorphy,
 we take advantage of Na$\ddot{\i}$ve Dimensional Analysis (NDA) \cite{nda}.
Before using NDA, we note that the terms in $W_{tree}$ are exactly
 proportional to $m_{1}/\Lambda_{H}$, $m_{3}/\Lambda_{H}$, $m_{5}/\Lambda_{H}$
 because of holomorphy.
In NDA, it is assumed that the other couplings in the effective Lagrangian are $O(1)$ in unit of $\Lambda_{H}$
 and that the effective theory also becomes strongly coupled at the scale $\Lambda_{H}$.
Therefore the effective Lagrangian at $\Lambda_{H}$ is expressed as
\begin{eqnarray}
{\cal L}_{eff} &\simeq& \frac{1}{(4 \pi)^{2}} \ \left[ 
\ \int {\rm d}^{4} \theta \ \Lambda_{H}^{2} \
\hat{K} \left( \frac{M^\prime}{\Lambda_{H}^{2}}, \frac{D^{\alpha}}{\Lambda_{H}}, \frac{M^{\prime \, \dagger}}{\Lambda_{H}^{2}}, \frac{\bar{D}_{\dot{\alpha}}}{\Lambda_{H}} \right) \right. \nonumber
\\
&+& \left. \int {\rm d}^{2} \theta \ \Lambda_{H}^{3} \ \left\{ \
\hat{W} \left( \frac{M^\prime}{\Lambda_{H}^{2}}, \frac{D^{\alpha}}{\Lambda_{H}} \right)
\ + \ \frac{m_{1}}{\Lambda_{H}} \frac{M_{12}^\prime}{\Lambda_{H}^{2}}
\ + \ \frac{m_{3}}{\Lambda_{H}} \frac{M_{34}^\prime}{\Lambda_{H}^{2}}
\ + \ \frac{m_{5}}{\Lambda_{H}} \frac{M_{56}^\prime}{\Lambda_{H}^{2}} \ \right\}
\ + {\rm h.c.} \ \right] \ , \nonumber
\\
\end{eqnarray}
 where the SM gauge interactions are omitted.
We rewrite the theory in terms of canonically normalized composite fields, $M_{ij}$, which are given by
\begin{eqnarray}
 M_{ij} &\simeq& \frac{1}{4 \pi \Lambda_{H}} M^\prime_{ij} \ \ \ {\rm at \ the \ scale} \ \Lambda_{H} \ ,
\end{eqnarray}
 and obtain the following canonical effective superpotential below $\Lambda_{H}$ 
 expressed in terms of $M_{ij}$'s:
\begin{eqnarray}
W_{eff} &\simeq& - \lambda \ \epsilon^{ijklmn} \ M_{ij} M_{kl} M_{mn}
\ + \ \xi_{\Omega} M_{12} \ + \ \xi_{\Phi} M_{34} \ + \ \xi M_{56} \ ,
\end{eqnarray}
 where $\lambda, \ \xi_{\Omega}, \ \xi_{\Phi}, \ \xi$ satisfy at the scale $\Lambda_{H}$
\begin{eqnarray}
\lambda(\Lambda_{H}) &\simeq& 4 \pi \ ,
\\
\xi_{\Omega}(\Lambda_{H}) &\simeq& \frac{ m_{1} \Lambda_{H} }{ 4 \pi } \ ,
\ \ \ \xi_{\Phi}(\Lambda_{H}) \ \simeq \  \frac{ m_{3} \Lambda_{H} }{ 4 \pi } \ ,
\ \ \ \xi(\Lambda_{H}) \ \simeq \ \frac{ m_{5} \Lambda_{H} }{ 4 \pi } \ .
\end{eqnarray}
Below the scale $\Lambda_{H}$, the physical couplings that correspond to 
 $\lambda, \ \xi_{\Omega}, \ \xi_{\Phi}, \ \xi$ 
 are regulated by the following renormalization group equations:
\begin{eqnarray}
\mu \frac{ {\rm d} }{ {\rm d}\mu } \left( \frac{1}{ \lambda^{2} } \right) &\simeq& - \frac{9}{16 \pi^{2}} \ ,
\\
\mu \frac{ {\rm d} }{ {\rm d}\mu } \xi &\simeq& \frac{3}{32 \pi^{2}} \lambda^{2} \ \xi \ ,
\ \ \ \mu \frac{ {\rm d} }{ {\rm d}\mu } \xi_{\Phi} \ \simeq \ \frac{3}{32 \pi^{2}} \lambda^{2} \ \xi_{\Phi} \ ,
\ \ \ \mu \frac{ {\rm d} }{ {\rm d}\mu } \xi_{\Omega} \ \simeq \ \frac{3}{32 \pi^{2}} \lambda^{2} \ \xi_{\Omega} \ .
\end{eqnarray}
Figure 1 shows the renormalization group running of the physical coupling $\lambda$ from the scale $\Lambda_{H}$ to lower scales.
For example, if $\Lambda_{H} \simeq 10$ TeV, $\lambda$ at the scale $M_{Z}$ is $\sim 2$.
The runnings of $\xi_{\Omega}, \ \xi_{\Phi}, \ \xi$ are not so drastic and can be neglected.
\begin{figure}[htbp]
  \begin{center}
   \includegraphics[width=90mm]{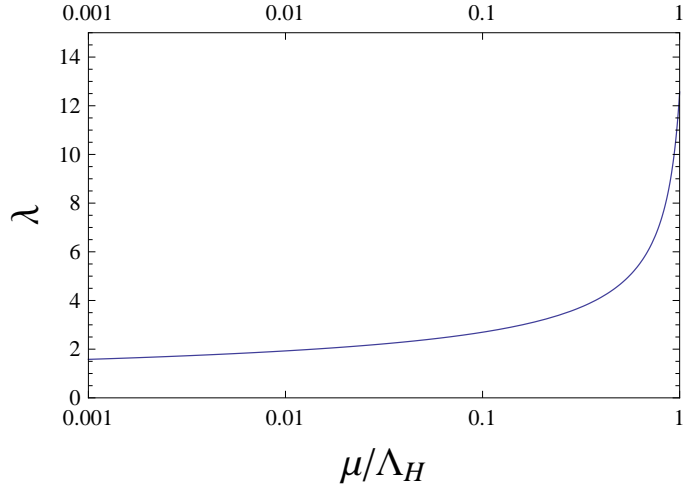}
  \end{center}
  \caption{The scale dependence of the physical coupling $\lambda$.
  }
\end{figure}

We rewrite the composite superfields $M_{ij}$ in the following way to clarify their SM charges:
\begin{eqnarray}
H_{u} &\equiv&
\left(
\begin{array}{c}
M_{13}  \\
M_{23}
\end{array}
\right) \ , \ \ \ 
H_{d} \ \equiv \
\left(
\begin{array}{c}
M_{14}  \\
M_{24}
\end{array}
\right) \ , \ \ \ 
\Phi_{u} \ \equiv \
\left(
\begin{array}{c}
M_{15}  \\
M_{25}
\end{array}
\right) \ , \ \ \ 
\Phi_{d} \ \equiv \
\left(
\begin{array}{c}
M_{16}  \\
M_{26}
\end{array}
\right) \ , \ \ \ \nonumber
\\
N &\equiv& M_{56} \ , \ \ \ 
N_{\Phi} \ \equiv \ M_{34} \ , \ \ \ 
N_{\Omega} \ \equiv \ M_{12} \ , \ \ \ \nonumber
\\
\Omega^{+} &\equiv& M_{35} \ , \ \ \ 
\Omega^{-} \ \equiv \ M_{46} \ , \ \ \ 
\zeta \ \equiv \ M_{36} \ , \ \ \ 
\eta \ \equiv \ M_{45} \ . \ \ \ 
\end{eqnarray}
Their SM charges and $Z_{2}$-parities are summarized in Table 2.
\begin{table}
\begin{center}
\begin{tabular}{|c|c|c|c|} \hline
Field                 & $SU(2)_{L}$ & $U(1)_{Y}$ & $Z_{2}$ \\ \hline
$H_{u}$               & 2           & +1/2       & +      \\ \hline
$H_{d}$               & 2           & $-$1/2       & +       \\ \hline
$\Phi_{u}$            & 2           & +1/2       & $-$       \\ \hline
$\Phi_{d}$            & 2           & $-$1/2       & $-$       \\ \hline
$\Omega^+$               & 1           & +1       & $-$      \\ \hline
$\Omega^-$               & 1           & $-$1       & $-$       \\ \hline
$N$, $N_{\Phi}$, $N_{\Omega}$               & 1           & 0       & +       \\ \hline
$\zeta$, $\eta$                             & 1           & 0       & $-$       \\ \hline
\end{tabular}
\end{center}
\caption{
Properties of the composite fields under the SM gauge groups and the $Z_2$ parity.
}
\end{table}
The effective superpotential is then written as
\begin{eqnarray}
W_{eff} &=& \lambda \ \left\{ \ N (H_{u}H_{d} + v_{0}^{2}) \ + \ N_{\Phi} (\Phi_{u}\Phi_{d} + v_{\Phi}^{2}) 
\ + \ N_{\Omega} (\Omega^{+}\Omega^{-} + v_{\Omega}^{2}) \right. \nonumber
\\
&-& \left.  N N_{\Phi} N_{\Omega} \ - \ N_{\Omega} \zeta \eta \ + \ 
\zeta H_{d} \Phi_{u} \ + \ \eta H_{u} \Phi_{d}
\ - \ \Omega^{+} H_{d} \Phi_{d} \ - \ \Omega^{-} H_{u} \Phi_{u} \ \right\} \ ,
\end{eqnarray}
 where $v_{0}^{2}, \ v_{\Phi}^{2}, \ v_{\Omega}^{2}$ are defined as
\begin{eqnarray}
v_{0}^{2} &\equiv& \xi/\lambda \ , \ \ \ v_{\Phi}^{2} \ \equiv \ \xi_{\Phi}/\lambda \ , \ \ \ 
v_{\Omega}^{2} \ \equiv \ \xi_{\Omega}/\lambda \ .
\end{eqnarray}
We note that all the three-point couplings are of the same magnitude in this model.
\\
\\

\section{Structure of the Effective Theory}

\ \ \ First of all, we look for vacua of the Higgs potential in the effective theory.
We study in the SUSY limit and then with soft SUSY breaking terms.
In the SUSY limit,
 the absolute minima of the superpotential (12) are determined from the tadpole conditions:
$(\partial/\partial \phi) W_{eff} = 0$ for each field $\phi$ at the tree level.
Since we are only interested in charge-conserving vacua,
 we set $H_{u}^{+}=H_{d}^{-}=\Phi_{u}^{+}=\Phi_{d}^{-}=\Omega^{+}=\Omega^{-}=0$
 and study whether the tadpole conditions can be satisfied.
Charge-conserving vacua are determined from the following tadpole conditions:
\begin{eqnarray}
0 &=& \frac{1}{\lambda} \frac{\partial W_{eff}}{\partial N_{\Omega}} \ = \ -N N_{\Phi} - \zeta \eta + v_{\Omega}^{2} \ ,
\\
0 &=& \frac{1}{\lambda} \frac{\partial W_{eff}}{\partial N} \ = \ -H_{u}^{0} H_{d}^{0} - N_{\Omega} N_{\Phi} + v_{0}^{2} \ ,
\\
0 &=& \frac{1}{\lambda} \frac{\partial W_{eff}}{\partial N_{\Phi}} \ = \ -\Phi_{u}^{0} \Phi_{d}^{0} - N_{\Omega} N + v_{\Phi}^{2} \ ,
\\
0 &=& \frac{1}{\lambda} \frac{\partial W_{eff}}{\partial \zeta} \ = \ H_{d}^{0} \Phi_{u}^{0} - N_{\Omega} \eta \ ,
\\
0 &=& \frac{1}{\lambda} \frac{\partial W_{eff}}{\partial \eta} \ = \ -H_{u}^{0} \Phi_{d}^{0} - N_{\Omega} \zeta \ ,
\\
0 &=& \frac{1}{\lambda} \frac{\partial W_{eff}}{\partial H_{u}^{0}} \ = \ -N H_{d}^{0} - \eta \Phi_{d}^{0} \ ,
\\
0 &=& \frac{1}{\lambda} \frac{\partial W_{eff}}{\partial \Phi_{u}^{0}} \ = \ -N_{\Phi} \Phi_{d}^{0} + \zeta H_{d}^{0} \ ,
\\
0 &=& \frac{1}{\lambda} \frac{\partial W_{eff}}{\partial H_{d}^{0}} \ = \ -N H_{u}^{0} + \zeta \Phi_{u}^{0} \ ,
\\
0 &=& \frac{1}{\lambda} \frac{\partial W_{eff}}{\partial \Phi_{d}^{0}} \ = \ -N_{\Phi} \Phi_{u}^{0} - \eta H_{u}^{0} \ ,
\\
0 &=& D \ = \ -\frac{1}{2} g_{1} \ ( H_{u}^{0 \, \dagger} H_{u}^{0} - H_{d}^{0 \, \dagger} H_{d}^{0} )
\ - \ \frac{1}{2} g_{1} \ ( \Phi_{u}^{0 \, \dagger} \Phi_{u}^{0} - \Phi_{d}^{0 \, \dagger} \Phi_{d}^{0} ) \ ,
\\
0 &=& D^{a=3} \ = \ -\frac{1}{2} g_{2} \ ( -H_{u}^{0 \, \dagger} H_{u}^{0} + H_{d}^{0 \, \dagger} H_{d}^{0} )
\ - \ \frac{1}{2} g_{2} \ ( -\Phi_{u}^{0 \, \dagger} \Phi_{u}^{0} + \Phi_{d}^{0 \, \dagger} \Phi_{d}^{0} ) \ ,
\end{eqnarray}
Since no symmetry forbids the term $m_{1} T_{1} T_{2}$ in the fundamental Lagrangian,
 we assume $v_{\Omega}^{2} \neq 0$.
Then the only solution to eqs. (14) and (19)-(22) is 
 $H_{d}^{0}=\Phi_{d}^{0}=H_{u}^{0}=\Phi_{u}^{0}=0$, i.e.
 the electroweak symmetry is unbroken in the absolute SUSY vacua.
At this point, our model distinctively differs from the minimal SUSY fat Higgs model \cite{fat},
 where non-anomalous $U(1)_{R}$ charges are assigned to forbid the term $m_{1} T_{1} T_{2}$
 so that the electroweak symmetry breaking does occur in the SUSY limit.
The D-terms are all zero in the absolute SUSY vacua because
 we have $H_{d}^{0}=\Phi_{d}^{0}=H_{u}^{0}=\Phi_{u}^{0}=0$.
The conditions in eqs.~(14)-(16) determine the VEVs of $N, N_{\Phi}$ and $N_{\Omega}$ as follows :
\begin{eqnarray}
\langle N \rangle \langle N_{\Phi} \rangle &=& v_{\Omega}^{2}, 
\ \ \ \langle N_{\Omega} \rangle \langle N_{\Phi} \rangle \ = \ v_{0}^{2}, 
\ \ \ \langle N_{\Omega} \rangle \langle N \rangle \ = \ v_{\Phi}^{2} \ .
\end{eqnarray}
We assume $v_{0}^{2} \neq 0$ and $v_{\Phi}^{2} \neq 0$
 as no symmetry forbids these terms.
We then have $\langle N_{\Omega} \rangle \neq 0$, which leads to $\eta=\zeta=0$ through eqs.~(17) and (18).
Note that the $Z_{2}$ parity is unbroken in the absolute SUSY vacua.

Since the conditions in eq. (25) have only one solution and the other neutral components are derived to be zero,
 we conclude that there is only one charge-conserving absolute SUSY vacuum 
 provided $v_{0}^{2}, \ v_{\Phi}^{2}$ and $v_{\Omega}^{2}$ are all non-zero.
This vacuum respects the electroweak symmetry and the $Z_{2}$ parity.
The non-zero VEVs of $N, \ N_{\Phi}$ and $N_{\Omega}$ give rise to effective $\mu$-terms.
\\

Let us proceed to the case with soft SUSY breaking terms.
For simplicity, we only introduce soft SUSY breaking mass terms and $B\mu$-term for $H_{u}$ and $H_{d}$,
 which we denote by $m_{H_{u}}^{2}, \ m_{H_{d}}^{2}$ and $B\mu$.
We redefine the phases of $T_{3}$ and $T_{5}$ to make $B \mu$ and $v_{0}^{2}$ real and positive.
We further rotate the phase of $(T_{1}, T_{2})$ so that the product of the VEVs of $N_{\Phi}$ and $N_{\Omega}$ is real.
We expand the potential with respect to $H_{u}, H_{d}, N, N_{\Omega}$ and $N_{\Phi}$,
 with setting $\Phi_{u}=\Phi_{d}=\Omega^{+}=\Omega^{-}=\zeta=\eta=0$.
The tree-level potential is then expressed as
\begin{eqnarray}
V &=& m_{H_{u}}^{2} H_{u}^{\dagger} H_{u} \ + \ m_{H_{d}}^{2} H_{d}^{\dagger} H_{d} 
\ + \ B\mu H_{u} H_{d} \ + \ h.c. \nonumber
\\
&+& \vert \lambda \vert^{2} \vert N \vert^{2} ( \vert H_{u} \vert^{2} + \vert H_{d} \vert^{2} ) \nonumber
\\
&+& \vert \lambda \vert^{2} \vert H_{u} H_{d} - N_{\Phi} N_{\Omega} + v_{0}^{2} \vert^{2} \nonumber
\\
&+& \vert \lambda \vert^{2} \vert N N_{\Omega} - v_{\Phi}^{2} \vert^{2} 
\ + \ \vert \lambda \vert^{2} \vert N N_{\Phi} - v_{\Omega}^{2} \vert^{2} \nonumber
\\
&+& \frac{1}{8} g_{1}^{2} \ ( H_{u}^{\dagger} H_{u} - H_{d}^{\dagger} H_{d} )^{2}
\ + \ \frac{1}{8} g_{2}^{2} \ ( H_{u}^{\dagger} \sigma^{a} H_{u} + H_{d}^{\dagger} \sigma^{a} H_{d} )^{2} \ .
\end{eqnarray}
Using the $SU(2)_{L}$ gauge symmetry, we take $H_{u}^{+} = 0$.
Then the condition: $( \partial/\partial H_{u}^{+} ) V = 0$ leads to $H_{d}^{-}=0$,
 as in the MSSM.
From the conditions:
 $( \partial/\partial H_{u}^{0 \, *} ) V = ( \partial/\partial H_{d}^{0 \, *} ) V = 
 ( \partial/\partial H_{u}^{0} ) V = ( \partial/\partial H_{d}^{0} ) V = 0$,
 we have
\begin{eqnarray}
\frac{1}{4} (g_{1}^{2} + g_{2}^{2}) ( \vert H_{u}^{0} \vert^{2} - \vert H_{d}^{0} \vert^{2}) H_{u}^{0}
&+& ( m_{H_{u}}^{2} + \vert \lambda \vert^{2} \vert N \vert^{2} + \vert \lambda \vert^{2} \vert H_{d}^{0} \vert^{2} ) H_{u}^{0}
\nonumber \\
&+&  ( \vert \lambda \vert^{2} N_{\Phi}N_{\Omega} - \vert \lambda \vert^{2} v_{0}^{2} - B\mu ) H_{d}^{0 \, *} \ = \ 0 \ ,
\\
- \frac{1}{4} (g_{1}^{2} + g_{2}^{2}) ( \vert H_{u}^{0} \vert^{2} - \vert H_{d}^{0} \vert^{2}) H_{u}^{0}
&+& ( m_{H_{d}}^{2} + \vert \lambda \vert^{2} \vert N \vert^{2} + \vert \lambda \vert^{2} \vert H_{u}^{0} \vert^{2} ) H_{d}^{0}
\nonumber \\
&+& ( \vert \lambda \vert^{2} N_{\Phi}N_{\Omega} - \vert \lambda \vert^{2} v_{0}^{2} - B\mu ) H_{u}^{0 \, *} \ = \ 0 \ ,
\\
H_{u}^{0}H_{d}^{0} &=& {\rm real} \ ,
\end{eqnarray}
 by using the fact that $B \mu$, $v_{0}^{2}$ and the VEV of $N_{\Phi} N_{\Omega}$ are real.
Since the VEV of $H_{u}^{0}H_{d}^{0}$ is real, we can take the VEVs of $H_{u}^{0}$ and $H_{d}^{0}$
 both real by using the $U(1)_{Y}$ gauge symmetry.
We hereafter denote these VEVs by $v_{u}$ and $v_{d}$, respectively.
The conditions: $( \partial/\partial N ) V = ( \partial/\partial N_{\Phi} ) V = 
 ( \partial/\partial N_{\Omega} ) V = 0$ and their complex conjugates lead to
\begin{eqnarray}
N(v_{u}^{2} + v_{d}^{2}) \ + \ N_{\Omega}^{*} (N N_{\Omega} - v_{\Phi}^{2}) 
 \ + \ N_{\Phi}^{*} (N N_{\Phi} - v_{\Omega}^{2}) &=& 0 \ ,
\\
N_{\Omega}^{*} (N_{\Phi} N_{\Omega} + v_{u}v_{d} - v_{0}^{2})
 \ + \ N^{*} (N N_{\Phi} - v_{\Omega}^{2}) &=& 0 \ ,
\\
N_{\Phi}^{*} (N_{\Phi} N_{\Omega} + v_{u}v_{d} - v_{0}^{2})
 \ + \ N^{*} (N N_{\Omega} - v_{\Phi}^{2}) &=& 0 \ ,
\\
N^{*} N_{\Phi}^{*} v_{\Omega}^{2}, \ N^{*} N_{\Omega}^{*} v_{\Phi}^{2} &=& {\rm real} \ .
\end{eqnarray}
\
\\

We derive the mass spectrum in the presence of soft SUSY breaking terms.
For simplicity, we here assume that $v_{\Phi}^{2}$ and $v_{\Omega}^{2}$ are also real and positive.
We assume that $\vert v_{u} \vert, \vert v_{d} \vert \ll \sqrt{v_{0}^{2}}, \sqrt{v_{\Phi}^{2}}, \sqrt{v_{\Omega}^{2}}$
 and we make a perturbative expansion of the masses with respect to $v_{u}^{2}$ and $v_{d}^{2}$.
At the zeroth order of $v_{u}^{2}$ and $v_{d}^{2}$,
 the VEVs of $N, N_{\Phi}$ and $N_{\Omega}$, 
 denoted by $\langle N \rangle^{0}, \ \langle N_{\Phi} \rangle^{0}$ and $\langle N_{\Omega} \rangle^{0}$,
 are given by
\begin{eqnarray}
\langle N \rangle^{0} &=& \sqrt{ \frac{v_{\Phi}^{2} v_{\Omega}^{2}}{v_{0}^{2}} } \ , \ \ \ 
\langle N_{\Phi} \rangle^{0} \ = \ \sqrt{ \frac{v_{0}^{2} v_{\Omega}^{2}}{v_{\Phi}^{2}} } \ , \ \ \ 
\langle N_{\Omega} \rangle^{0} \ = \ \sqrt{ \frac{v_{0}^{2} v_{\Phi}^{2}}{v_{\Omega}^{2}} } \ ,
\end{eqnarray}
 which are the same as those in the SUSY limit.
These VEVs respectively correspond to the SUSY-conserving masses of 
 $(H_{u}, H_{d})$, $(\Phi_{u}, \Phi_{d})$ and $(\Omega^{+}, \Omega^{-})$.
In the following discussion, we use the VEVs of $N^{0}, N_{\Phi}^{0}$ and $N_{\Omega}^{0}$
 as the parameters of the model, instead of $v_{0}^{2}, v_{\Phi}^{2}$ and $v_{\Omega}^{2}$.
The VEVs of $N, N_{\Phi}$ and $N_{\Omega}$ at the first order of $v_{u}^{2}$ and $v_{d}^{2}$,
 denoted by $\langle N \rangle^{1}, \langle N_{\Phi} \rangle^{1}$ and $\langle N_{\Omega} \rangle^{1}$, 
 satisfy the following relations:
\begin{eqnarray}
& & \left(
\begin{array}{ccc}
(\langle N_{\Phi} \rangle^{0})^{2} + (\langle N_{\Omega} \rangle^{0})^{2} & \langle N \rangle^{0} \langle N_{\Phi} \rangle^{0} & \langle N \rangle^{0} \langle N_{\Omega} \rangle^{0} \\
 & (\langle N \rangle^{0})^{2} + (\langle N_{\Omega} \rangle^{0})^{2} & \langle N_{\Phi} \rangle^{0} \langle N_{\Omega} \rangle^{0} \\
 & & (\langle N \rangle^{0})^{2} + (\langle N_{\Phi} \rangle^{0})^{2}
\end{array}
\right)
\left(
\begin{array}{c}
\langle N \rangle^{1}  \\
\langle N_{\Phi} \rangle^{1}  \\
\langle N_{\Omega} \rangle^{1}
\end{array}
\right) \nonumber
\\
 &=& 
\left(
\begin{array}{c}
-\langle N \rangle^{0} (v_{u}^{2}+v_{d}^{2}) \\
0  \\
0 
\end{array}
\right) \ .
\end{eqnarray}
We next study the mass spectrum of the $Z_{2}$-even Higgs sector.
At the first order of $v_{u}^{2}$ and $v_{d}^{2}$,
 the conditions (26) and (27) reduce to
\begin{eqnarray}
\frac{1}{4} (g_{1}^{2} + g_{2}^{2}) (v_{u}^{2} - v_{d}^{2}) v_{u}
&+& ( \ m_{H_{u}}^{2} + \vert \lambda \vert^{2} \vert \langle N \rangle^{0} + \langle N \rangle^{1} \vert^{2} + \vert \lambda \vert^{2} v_{d}^{2} \ ) \ v_{u}
\nonumber
\\
&+& \{ \ \vert \lambda \vert^{2} \ ( \ \langle N_{\Phi} \rangle^{1} \langle N_{\Omega} \rangle^{0} + \langle N_{\Phi} \rangle^{0} \langle N_{\Omega} \rangle^{1} \ ) \ - \ B\mu \ \} \ v_{d} \ = \ 0 \ ,
\nonumber
\\
\\
- \frac{1}{4} (g_{1}^{2} + g_{2}^{2}) (v_{u}^{2} - v_{d}^{2}) v_{d}
&+& ( \ m_{H_{d}}^{2} + \vert \lambda \vert^{2} \vert \langle N \rangle^{0} + \langle N \rangle^{1} \vert^{2} + \vert \lambda \vert^{2} v_{u}^{2} \ ) \ v_{d}
\nonumber
\\
&+& \{ \ \vert \lambda \vert^{2} \ ( \ \langle N_{\Phi} \rangle^{1} \langle N_{\Omega} \rangle^{0}  + \langle N_{\Phi} \rangle^{0} \langle N_{\Omega} \rangle^{1} \ ) \ - \ B\mu \ \} \ v_{u} \ = \ 0 \ .
\nonumber
\\
\end{eqnarray}
They give the same conditions for the electroweak symmetry breaking as 
 in the Next-to-MSSM~\cite{NMSSM}
 if we define the effective $B\mu$-term as 
\begin{eqnarray}
B \mu_{eff} &\equiv& B \mu 
\ - \ \vert \lambda \vert^{2} \ ( \ \langle N_{\Phi} \rangle^{1} \langle N_{\Omega} \rangle^{0}  + \langle N_{\Phi} \rangle^{0} \langle N_{\Omega} \rangle^{1} \ ) \ .
\end{eqnarray}
\\

We comment on the range of $\Lambda_{H}$ favored by the naturalness.
NDA implies that the Lagrangian contains the following K\"ahler potential:
\begin{eqnarray}
{\cal L}_{eff} &\supset& \int {\rm d}^{4} \theta \ \frac{\Lambda_{H}}{4\pi} \ ( \ N + N_{\Phi} + N_{\Omega} \ ) \ 
\end{eqnarray}
 with a $O(1)$ factor for each coupling.
We now turn on the effects of soft SUSY breaking. 
We introduce a singlet chiral superfield $X$ whose F-term, $F_{X}$, has a non-zero VEV.
We couple $X$ to the other superfields
 by contact interactions suppressed by the scale $M$, where
 $M$ is indentified with the Planck scale in the case of gravity mediation
 and with the messenger scale in the case of gauge mediation
\footnote{Notice that since the superfields $N, N_{\Phi}$ and $N_{\Omega}$ are composites of the fundamental superfields $T_{i}$'s,
 which are charged under the $SU(2)_{H}$ gauge group,
 gauge mediation can induce soft SUSY breaking terms for $N, N_{\Phi}$ and $N_{\Omega}$ even if they are gauge singlets.}.
The soft SUSY breaking scale is given by $F_{X}/M$.
With soft SUSY breaking effects,
 we have the following K\"ahler potential that gives extra contributions to the tadpole terms in eq. (6):
\begin{eqnarray}
{\cal L}_{eff} &\supset& \int {\rm d}^{4} \theta \ 
\frac{X^{\dagger}}{M} \ \frac{\Lambda_{H}}{4\pi} \ ( \ N + N_{\Phi} + N_{\Omega} \ ) \ + \ {\rm h.c.} \nonumber
\\
&=& \frac{F_{X}^{\dagger}}{M} \ \int {\rm d}^{2} \theta \ 
\frac{\Lambda_{H}}{4\pi} \ ( \ N + N_{\Phi} + N_{\Omega} \ ) \ + \ {\rm h.c.} \ .
\end{eqnarray}
These extra contributions affect the VEV of $N$, which is given by
\begin{eqnarray}
\langle N \rangle &=& \sqrt{ \frac{1}{\lambda(M_{Z})} } \sqrt{ \frac{ \xi^{\prime}_{\Phi} \ \xi^{\prime}_{\Omega} }{\xi^{\prime}} } \ ,
\end{eqnarray}
 where $\xi^{\prime}, \xi^{\prime}_{\Phi}$ and $\xi^{\prime}_{\Omega}$ are the sums of the tadpoles in eq. (6)
 and the extra contributions from soft SUSY breaking effects, which are expressed as
\begin{eqnarray}
\xi^{\prime} &=& \xi \ + \ a \frac{F_{X}^{\dagger}}{M} \frac{\Lambda_{H}}{4\pi} \ , 
\ \ \ 
\xi_{\Phi}^{\prime} \ = \ \xi_{\Phi} \ + \ a_{\Phi} \frac{F_{X}^{\dagger}}{M} \frac{\Lambda_{H}}{4\pi} \ , 
\ \ \ 
\xi_{\Omega}^{\prime} \ = \ \xi_{\Omega} \ + \ a_{\Omega} \frac{F_{X}^{\dagger}}{M} \frac{\Lambda_{H}}{4\pi} \ ,
\end{eqnarray}
 where $a, a_{\Phi}$ and $a_{\Omega}$ are all $O(1)$.
On the other hand, 
 to break the electroweak symmetry without unnatural cancellation between 
 the effective $\mu$-term $\lambda(M_{Z}) \langle N \rangle$ and 
 the soft SUSY breaking terms $m_{H_{u}}^{2}$ and $m_{H_{d}}^{2}$,
 we require that $\lambda(M_{Z}) \langle N \rangle$ be smaller than about 1 TeV,
 which is equivalent to
\begin{eqnarray}
\sqrt{ \frac{ \xi^{\prime}_{\Phi} \ \xi^{\prime}_{\Omega} }{\xi^{\prime}} } 
&\lesssim& \frac{1}{\sqrt{\lambda(M_{Z})}} \ \times \ 1 \ {\rm TeV} \ .
\end{eqnarray}
To satisfy the above condition in the presence of soft SUSY breaking effects,
 it is required that
\begin{eqnarray}
\sqrt{ \left\vert \frac{F_{X}}{M} \right\vert \frac{\Lambda_{H}}{4\pi} } &\lesssim& \frac{1}{\sqrt{\lambda(M_{Z})}} \ \times \ 1 \ {\rm TeV} \ .
\end{eqnarray}
Assuming that the soft SUSY breaking scale $F_{X}/M$ is about 1 TeV and $\lambda(M_{Z})$ is $O(1)$,
 we obtain the following estimate on the upper bound of $\Lambda_{H}$ suggested by the naturalness:
\begin{eqnarray}
\Lambda_{H} &\lesssim& 4 \pi \ \times \ 1 \ {\rm TeV} \ .
\end{eqnarray}
Now that we know the favorable range of $\Lambda_{H}$,
 we discuss the mass of the SM-like Higgs boson, $m_h$, which depends on $\Lambda_{H}$ via $\lambda(M_{Z})$.
In the decoupling limit,
 $m_{h}$ is given at the tree level by~\cite{NMSSM}
\begin{eqnarray}
m_{h}^{2} &\simeq& M_{Z}^{2} \cos^{2}(2\beta) \ + \ \lambda(M_{Z})^{2} v^{2} \sin^{2}(2 \beta) \ .
\end{eqnarray}
Here we introduce $\tan\beta \equiv v_u/v_d$ as usual.
Since the naturalness suggests $\Lambda_{H} \lesssim 10$ TeV, 
 we have $\lambda(M_{Z}) > 1.9$, according to Figure 1.
Therefore, to realize the Higgs boson mass of $125$ GeV,
 we need large $\tan \beta$ so that the contribution from the second term in eq. (46)
 is suppressed.
\\

Finally we comment on how the SM Yukawa couplings of appropriate magnitudes
 can be derived in this model, especially for the $O(1)$ top quark Yukawa coupling.
We adopt the mechanism proposed in refs. \cite{techful, fat}.
First we introduce two additional $SU(2)_{H}$ doublets, $T_{7}$ and $T_{8}$, which have 
 the tree-level mass term below:
\begin{eqnarray}
W_{7} &=& m_{7} T_{7}T_{8} \ ,
\end{eqnarray}
 where $m_{7} > \Lambda_{H}$ is assumed.
Above the scale $m_{7}$, our model is described by the SUSY QCD theory with two colors and four flavors
 and is in the superconformal window.
We assume that
 this theory approaches the infrared fixed point at some scale $\Lambda_{4} > m_{7}$
 and becomes nearly conformal.
At the scale $m_{7}$, the fields $T_{7}$ and $T_{8}$ decouple and the theory becomes 
 the SUSY QCD theory with two colors and three flavors.
Through renormalization group evolutions, the wavefunctions of $T_i$'s receive large corrections of 
\begin{eqnarray}
Z &\simeq& \left( \frac{ m_{7} }{ \Lambda_{4} } \right)^{\gamma^{*}} \ ,
\end{eqnarray}
 where $\gamma^{*}$ denotes the anomalous dimension at the infrared fixed point,
 which equals to $(3N_{c}-N_{f})/N_{f} = 1/2$
 for the $N_{c}=2$ and $N_{f}=4$ SUSY QCD theory.
By introducing new $SU(2)_{H}$ singlets with SUSY-conserving mass $M_{f} \sim \Lambda_{H}$ and integrating them out,
 it is possible to have the following higher-dimensional superpotential at the scale $\Lambda_{H}$:
\begin{eqnarray}
W_{f} &=& \frac{1}{M_{f}} Z^{-1} h_{u}^{ij} (T_{1}T_{3}, T_{2}T_{3}) Q_{i} U_{j} \ + \ \cdots ,
\end{eqnarray}
 where $h_{u}^{ij}$ denotes the Yukawa coupling in the fundamental theory, whose value is at most $O(1)$
 and the factor of $Z^{-1}$ comes from the wavefunction renormalization of $T_{i}$'s.
Therefore the fundamental theory may not contain any Landau pole
 below the Planck scale, and the theory can be UV-complete.
We find that, below the scale $\Lambda_{H}$, the term in eq. (49) reduces to the following term:
\begin{eqnarray}
W_{f} &\simeq& \frac{\Lambda_{H}}{4 \pi M_{f}} Z^{-1} h_{u}^{ij} H_{u} Q_{i} U_{j} \ + \ \cdots
\end{eqnarray}
 by using NDA.
We can cancel the suppression factor of $1/4\pi$ by the enhancement factor of $Z^{-1}$
 with appropriate values of $\Lambda_{4}$ and $m_{7}$. 
In particular, we can derive the $O(1)$ top quark Yukawa coupling below $\Lambda_H$.
\\
\\

\section{A Natural Model for the Light Higgs Boson
from SUSY Strong Dynamics}

\ \ \ In this section, we present a simple UV-complete model where
 the SM-like Higgs boson strongly couples to other fields in the Higgs sector,
 but its mass is naturally as light as 125 GeV.
This model arises as the low-energy effective theory of the SUSY QCD theory 
 with $N_c=2$ and $N_f=3$ with an additional
 $SU(2)_{H}$ singlet chiral superfield.
The singlet induces a large SUSY-conserving mass for $N$, so that
 $N$ decouples from the theory below $\Lambda_{H}$ and
 the term $N H_{u} H_{d}$ disappears from the effective superpotential.
The effective theory has a similar structure to
 the ``Four Higgs doublets and two charged singlets" (4HD$\Omega$) model \cite{4hdo}.
\\

We introduce a $SU(2)_{H}$ singlet chiral superfield, $S$, which is neutral under the SM gauge groups
 and is $Z_{2}$-even.
The superpotential involving $S$ generally takes the following form 
\footnote{The possible tadpole term for $S$ can be eliminated
 by shifting the value of $S$.}:
\begin{eqnarray}
\Delta W &=& ( \ y_{1} T_{1}T_{2} \ + \ y_{3} T_{3}T_{4} \ + \ y_{5} T_{5}T_{6} \ ) \ S \ 
+ \ \frac{M_{S}}{2} S^{2} \ + \ \frac{\kappa}{3} S^{3} \ .
\end{eqnarray}
The coupling constants $y_{1}, y_{3}$ and $y_{5}$ are assumed to be at most $O(1)$ at the scale $\Lambda_{4}$,
 so that they remain finite up to the Planck scale.
For simplicity, we here assume $y_{5} \gg y_{1}, y_{3}$
\footnote{When $y_{1}$ and $y_{3}$ are as large as $y_{5}$, we have to take into account
 the mixings among $N, N_{\Phi}$ and $N_{\Omega}$ in the effective theory.
This complicates the model, although it does not affect the main results of our discussion.
}.
At scales below $\Lambda_{4}$, they are enhanced by the same mechanism as the Yukawa couplings described 
 in the previous section;
 renormalization group running from the scale $\Lambda_{4}$ to $m_{7}$ 
 enhances them by the factor of $Z^{-1}$.
We assume that $M_{S}$ is of the order of $\Lambda_{H}$.
Then $S$ can be integrated out in the effective theory below $\Lambda_{H}$.
We obtain
\begin{eqnarray}
\Delta W &=& - \frac{y_{5}^{2} Z^{-2}}{2 M_{S}} \ (T_{5}T_{6}) (T_{5}T_{6}) \ .
\end{eqnarray}
After using NDA, this becomes the following term
 in the effective superpotential:
\begin{eqnarray}
\Delta W_{eff} &\simeq& - \frac{ y_{5}^{2}Z^{-2} }{2 M_{S}} \frac{\Lambda_{H}^{2}}{(4 \pi)^{2}} \ N^{2}
\ = \ -\frac{M_{N}}{2} N^{2}
\end{eqnarray}
 where $M_N$ is defined as
\begin{eqnarray}
M_{N} &\equiv& \frac{y_{5}^{2} Z^{-2}}{M_{S}} \frac{\Lambda_{H}^{2}}{(4 \pi)^{2}} \ .
\end{eqnarray}
Since the factor of $Z^{-2}$ compensates the suppression factor of $1/(4 \pi)^{2}$,
 we have the relation: $M_{N} \sim \Lambda_{H}$ when $y_{5}$ is $O(1)$.
\\

We study how the term in eq. (53)  
 modifies the model with the relation $M_{N} \sim \Lambda_{H}$.
First we look for charge-conserving absolute SUSY vacua,
 using the conditions eqs. (14)-(24), but eq. (15) is replaced with
\begin{eqnarray}
0 &=& -H_{u}^{0}H_{d}^{0} - N_{\Omega}N_{\Phi} + v_{0}^{2} - \frac{M_{N}}{\lambda} N \ .
\end{eqnarray}
From eqs. (14) and (19)-(22), we again have $H_{d}^{0}=\Phi_{d}^{0}=H_{u}^{0}=\Phi_{u}^{0}=0$.
It follows that $\zeta=\eta=0$.
We obtain the following relations:
\begin{eqnarray}
N N_{\Phi} &=& v_{\Omega}^{2}, \ \ \ N_{\Omega} N_{\Phi} \ = \ v_{0}^{2} - \frac{M_{N}}{\lambda} N, \ \ \ N_{\Omega} N \ = \ v_{\Phi}^{2} \ .
\end{eqnarray}
Let us take those values of $m_{1}, m_{3}$ and $m_{5}$ which satisfy
 the following relations: 
\begin{eqnarray}
m_{5}^{2} &\gtrsim& 4 \pi \Lambda_{H} m_{1} / \lambda(M_{Z}) \ , 
\ \ \ m_{5}^{2} \ \gtrsim \ 4 \pi \Lambda_{H} m_{3} / \lambda(M_{Z}) \ ,
\\
\lambda(M_{Z}) \frac{m_{5} \Lambda_{H}}{4 \pi} &\ll& \Lambda_{H}^{2} \ ,
\end{eqnarray}
 or equivalently
\begin{eqnarray}
\xi^{2} &\gtrsim& \xi_{\Phi} M_{N}^{2} / \lambda(M_{Z}) \ , 
\ \ \ \xi^{2} \ \gtrsim \ \xi_{\Omega} M_{N}^{2} / \lambda(M_{Z}) \ ,
\\
\lambda(M_{Z}) \xi &\ll& M_{N}^{2} \ ,
\end{eqnarray}
 where the scale dependences of $\xi, \xi_{\Phi}$ and $\xi_{\Omega}$ are negligible.
The VEV of $N$ is then approximately given by
\begin{eqnarray}
\langle N \rangle &\simeq& \frac{\xi}{M_{N}} \ = \ \frac{\lambda v_{0}^{2}}{M_{N}} \ ,
\end{eqnarray}
 and eventually the VEVs of $N_{\Phi}$ and $N_{\Omega}$ are given by
\begin{eqnarray}
\langle N_{\Phi} \rangle \ \simeq \ \frac{1}{\lambda} \frac{\xi_{\Omega}}{\xi} M_{N} \ = \ 
\frac{1}{\lambda} \frac{v_{\Omega}^{2}}{v_{0}^{2}} M_{N} \ , \ \ \ 
\langle N_{\Omega} \rangle \ \simeq \ \frac{1}{\lambda} \frac{\xi_{\Phi}}{\xi} M_{N} \ = \
\frac{1}{\lambda} \frac{v_{\Phi}^{2}}{v_{0}^{2}} M_{N} \ .
\end{eqnarray}
Since $N$ has the large SUSY-conserving mass of order $\Lambda_{H}$,
 we may integrate it out in the effective theory below $\Lambda_{H}$.
We then obtain the following superpotential:
\begin{eqnarray}
W_{eff} &=& \lambda \ \left\{ \ N_{\Phi} (\Phi_{u}\Phi_{d} + v_{\Phi}^{2}) 
\ + \ N_{\Omega} (\Omega^{+}\Omega^{-} + v_{\Omega}^{2}) \right. \nonumber
\\
&-& \left. N_{\Omega} \zeta \eta \ + \ 
\zeta H_{d} \Phi_{u} \ + \ \eta H_{u} \Phi_{d}
\ - \ \Omega^{+} H_{d} \Phi_{d} \ - \ \Omega^{-} H_{u} \Phi_{u} \ \right\} \nonumber
\\
&+& \frac{\lambda^{2}}{2M_{N}} ( H_{u}H_{d} + v_{0}^{2} - N_{\Phi}N_{\Omega} )^{2} \ .
\end{eqnarray}
We expand the fields $N_{\Phi}$ and $N_{\Omega}$ around their VEVs
 and replace them respectively with $\langle N_{\Phi} \rangle+n_{\Phi}$ and $\langle N_{\Omega} \rangle+n_{\Omega}$,
 where $n_{\Phi}$ and $n_{\Omega}$ denote their physical components.
The superpotential is then expressed as
\begin{eqnarray}
W_{eff} &=& \lambda \ \left\{ \ \left( \frac{v_{\Omega}^{2}}{\lambda v_{0}^{2}}M_{N} + n_{\Phi} \right) (\Phi_{u}\Phi_{d} + v_{\Phi}^{2}) 
\ + \ \left( \frac{v_{\Phi}^{2}}{\lambda v_{0}^{2}}M_{N} + n_{\Omega} \right) (\Omega^{+}\Omega^{-} + v_{\Omega}^{2}) \right. \nonumber
\\
&-& \left. \left( \frac{v_{\Phi}^{2}}{\lambda v_{0}^{2}}M_{N} + n_{\Omega} \right) \zeta \eta \ + \ 
\zeta H_{d} \Phi_{u} \ + \ \eta H_{u} \Phi_{d}
\ - \ \Omega^{+} H_{d} \Phi_{d} \ - \ \Omega^{-} H_{u} \Phi_{u} \ \right\} \nonumber
\\
&+& \frac{\lambda^{2}}{2M_{N}} \left( H_{u}H_{d} + v_{0}^{2} - \frac{v_{\Phi}^{2}v_{\Omega}^{2}}{\lambda^{2} v_{0}^{4}} M_{N}^{2}
\ - \ \frac{v_{\Omega}^{2}}{\lambda v_{0}^{2}}M_{N} n_{\Omega} 
\ - \ \frac{v_{\Phi}^{2}}{\lambda v_{0}^{2}}M_{N} n_{\Phi}
\ - \ n_{\Phi}n_{\Omega}
\right)^{2} \ ,
\end{eqnarray}
 which can be rewritten as
\begin{eqnarray}
	W_{eff} &=&
	-\mu (H_uH_d-n_{\Phi}n_{\Omega})-\mu_{\Phi}\Phi_u\Phi_d-\mu_{\Omega}(\Omega^+\Omega^- - \zeta\eta)
	\nonumber\\
	&+&\lambda \left\{ \
	H_d\Phi_u\zeta + H_u\Phi_d\eta - H_u\Phi_u\Omega^- - H_d\Phi_d\Omega^+
	+n_{\Phi}\Phi_u\Phi_d +n_{\Omega}(\Omega^+\Omega^- -\zeta\eta) \ \right\}
	\nonumber\\
	&+& \cdots \ ,
\end{eqnarray}
 where the effective $\mu$-terms are given as
\begin{eqnarray}
\mu &=& -\frac{\lambda^{2}}{M_N} \left( v_{0}^{2} - \frac{v_{\Phi}^{2}v_{\Omega}^{2}}{\lambda^{2} v_{0}^{4}}M_{N}^{2} \right)
\ \simeq \ -\frac{\lambda^{2} v_0^{2}}{M_N} \ , \ \ \
\mu_{\Phi} \ = \ -\frac{v_{\Omega}^{2}}{v_{0}^{2}} M_N \ , \ \ \
\mu_{\Omega} \ = \ -\frac{v_{\Phi}^{2}}{v_{0}^{2}} M_N \ .
\end{eqnarray}
For $\mu$, we make an approximation using the relations in eqs. (59) and (60).
Notice that we have $\vert \mu \vert \gtrsim \vert \mu_{\Phi} \vert, \ 
 \vert \mu \vert \gtrsim \vert \mu_{\Omega} \vert$ because of the relation in eq. (59).

So far, we have shown that 
\begin{itemize}
 \item $N$ can be integrated out from the effective theory below $\Lambda_{H}$
  so that there is no three-point coupling for $H_{u}$ and $H_{d}$.
 \item The VEV of $N$ still gives the effective $\mu$-term, which can take an appropriate value.
 \item The VEVs of $N_{\Phi}$ and $N_{\Omega}$ can be of the same order as or smaller than that of $N$
  so that $\Phi_{u,d}$ and $\Omega^{\pm}$ do not decouple.
\end{itemize}
Consequently, the model can reduce to
 the ``Four Higgs doublets and two charged singlets" (4HD$\Omega$) model \cite{4hdo},
 with additional superfields $n_{\Phi}$ and $n_{\Omega}$ which have little impact on
 the main features of the model because their tree-level couplings to the MSSM Higgs fields, $H_{u}$ and $H_{d}$,
 are suppressed by powers of $1/\Lambda_{H}$.
\\

Before discussing the mass of the SM-like Higgs boson,
 we again comment on the range of $\Lambda_{H}$ that is favored by the naturalness.
The naturalness of the electroweak symmetry breaking requires that 
 the effective $\mu$-term for $H_{u}$ and $H_{d}$ be smaller than about 1 TeV:
\begin{eqnarray}
\lambda(M_{Z}) \langle N \rangle \ = \ \mu &\lesssim& 1 \ {\rm TeV} \ .
\end{eqnarray}
Additionally, in order that the fields $\Phi_{u}, \Phi_{d}$ and $\Omega^{\pm}$
 remain in the effective theory, the terms of $\mu_{\Phi}$ and $\mu_{\Omega}$ should be
 smaller than 1 TeV:
\begin{eqnarray}
\lambda(M_{Z}) \langle N_{\Phi} \rangle \ = \ \mu_{\Phi} \ \lesssim \ 1 \ {\rm TeV} \ , \ \ \ \lambda(M_{Z}) \langle N_{\Omega} \rangle \ = \ \mu_{\Omega} &\lesssim& 1 \ {\rm TeV} \ .
\end{eqnarray}
On the other hand, soft SUSY breaking terms in the K\"ahler potential:
\begin{eqnarray}
{\cal L}_{eff}
&\supset& \frac{F_{X}^{\dagger}}{M} \ \int {\rm d}^{2} \theta \ 
\frac{\Lambda_{H}}{4\pi} \ ( \ N + N_{\Phi} + N_{\Omega} \ ) \ + \ {\rm h.c.} \ .
\end{eqnarray}
 give extra contributions to the tadpoles in eq. (5), as we have discussed in Section 3.
The VEVs of $N, N_{\Phi}$ and $N_{\Omega}$ are then given by
\begin{eqnarray}
\langle N \rangle &\simeq& \frac{\xi^{\prime}}{M_{N}} \ , \ \ \ 
\langle N_{\Phi} \rangle \ \simeq \ \frac{1}{\lambda(M_{Z})} \frac{\xi_{\Omega}^{\prime}}{\xi^{\prime}} M_{N} \ , \ \ \ 
\langle N_{\Omega} \rangle \ \simeq \ \frac{1}{\lambda(M_{Z})} \frac{\xi_{\Phi}^{\prime}}{\xi^{\prime}} M_{N} \ ,
\end{eqnarray}
 where $\xi^{\prime}, \xi^{\prime}_{\Phi}$ and $\xi^{\prime}_{\Omega}$ contain 
 the extra contributions from soft SUSY breaking effects and can be written as
\begin{eqnarray}
\xi^{\prime} &=& \xi \ + \ a \frac{F_{X}^{\dagger}}{M} \frac{\Lambda_{H}}{4\pi} \ , 
\ \ \ 
\xi_{\Phi}^{\prime} \ = \ \xi_{\Phi} \ + \ a_{\Phi} \frac{F_{X}^{\dagger}}{M} \frac{\Lambda_{H}}{4\pi} \ , 
\ \ \ 
\xi_{\Omega}^{\prime} \ = \ \xi_{\Omega} \ + \ a_{\Omega} \frac{F_{X}^{\dagger}}{M} \frac{\Lambda_{H}}{4\pi} \ ,
\end{eqnarray}
 where $a, a_{\Phi}$ and $a_{\Omega}$ are all $O(1)$.
In order to solve the gauge hierarchy problem,
 the soft SUSY breaking scale $F_{X}/M$ is at most 1 TeV.
Soft SUSY breaking then always respect the condition eq. (67).
To satisfy the conditions eqs. (68) in the presence of soft SUSY breaking effects,
 we need to have 
\begin{eqnarray}
\Lambda_{H} &\lesssim& 4 \pi \ \times \ 1 \ {\rm TeV} \ .
\end{eqnarray}

As we know the range of $\Lambda_{H}$ that is favored by the naturalness 
 and that gives not so heavy $\Phi_u, \Phi_d$ and $\Omega^{\pm}$,
 we calculate the mass of the SM-like Higgs boson $m_h$.
The mass depends on the coupling constant $\lambda$
 through radiative corrections involving the scalar and fermion components
 of $Z_2$-odd fields $\Omega^{\pm}$, $\Phi_{u}^{+}$, $\Phi_{d}^{-}$, $\zeta$ and $\eta$.
Hence $m_h$ depends on $\Lambda_{H}$ through the scale dependence of $\lambda$.
Here we consider the case with $\mu_{\Omega}=\mu_{\Phi}=0$ for simplicity.
The charginos still obtain masses from the VEVs of the Higgs fields.
In this simple case, the $Z_2$-odd scalars do not mix with each other.
Thus the corrected mass of the SM-like Higgs boson is approximately expressed as 
\begin{equation}
	m_h^2\simeq m_Z^2\cos^22\beta+(\text{MSSM-loop})
	+\frac{\lambda^4v^2}{8\pi^2}\left(
		c_{\beta}^4
		\ln\frac{m_{\Omega^+}^2m_{\Phi_d^{\pm}}^2m_{\Phi_u^{0}}^2m_{\zeta}^2}
		{m_{\tilde{\chi}_1^{\prime \pm}}^4m_{\tilde{\chi}_1^{\prime 0}}^4}
		+s_{\beta}^4\ln\frac{m_{\Omega^-}^2m_{\Phi_u^{\pm}}^2m_{\Phi_d^{0}}^2m_{\eta}^2}
		{m_{\tilde{\chi}_2^{\prime \pm}}^4m_{\tilde{\chi}_2^{\prime 0}}^4}
	\right)\;,
\end{equation}
where 
$m_{\Phi_{u}^{\pm}}(m_{\Phi_d^{\pm}})$ and $m_{\Phi_{u}^{0}} (m_{\Phi_d^0})$ 
are the masses of charged scalars and neutral scalars from $Z_2$-odd doublet $\Phi_u(\Phi_d)$,
$m_{\Omega^{\pm}}$ and $m_{\zeta,\eta}$ are the scalar masses of 
$Z_2$-odd charged singlets and $Z_2$-odd neutral singlets respectively, 
$m_{\tilde{\chi}_{1,2}^{\prime \pm}}$ are the $Z_2$-odd chargino masses, and
$m_{\tilde{\chi}_{1,2}^{\prime 0}}$ are the $Z_2$-odd neutralino masses.
The mass eigenstates of the neutralinos and charginos in this simple case are written as 
$\tilde{\chi}_1^{\prime 0}=(\tilde{\zeta},\bar{\tilde{\Phi}}_d^0)^T$,
$\tilde{\chi}_2^{\prime 0}=(\tilde{\eta},\bar{\tilde{\Phi}}_u^0)^T$,
$\tilde{\chi}_1^{\prime +}=(\tilde{\Omega}^+,\bar{\tilde{\Phi}}_d^+)^T$, and
$\tilde{\chi}_1^{\prime -}=(\tilde{\Omega}^-,\bar{\tilde{\Phi}}_u^-)^T$.
The $Z_2$-odd scalar masses can be typically expressed by 
$m_{\phi^{\prime}}^2=\bar{m}_{\phi^{\prime}}^2+(k^{\prime}g^{\prime 2}+k g^2+c\lambda^2)v^2$,
where $\bar{m}_{\phi^{\prime}}^2$ denotes the soft SUSY breaking scalar mass~\footnote{
	Because $\Phi_u^{\pm}$ ($\Phi_d^{\pm}$) and $\Phi_u^{0}$($\Phi_d^{0}$) are 
	the components of $\Phi_u$ ($\Phi_d$), 
	$\bar{m}_{\Phi_u^{\pm}}^2=\bar{m}_{\Phi_u^{0}}^2$
	and 
	$\bar{m}_{\Phi_d^{\pm}}^2=\bar{m}_{\Phi_d^{0}}^2$
	are satisfied.
}.
For the soft SUSY breaking masses, we consider three benchmark sets of parameters shown in 
 Table~\ref{paraset}.
When the contribution coming from the Higgs VEV dominates the mass, 
 significant non-decoupling effects can arise \cite{hhh1, hhh2}. 
Since we are interested in such non-decoupling cases, some soft SUSY breaking parameters
 are taken to be as light as 50\;GeV.
\begin{table}
	\caption{Benchmark sets of model parameters. 
	$\mu_{\Phi}$, $\mu_{\Omega}$ and the relevant B-terms are 
set to be zero for simplicity.
For the MSSM parameters, we fix 
$\bar{m}_{\tilde{t}_L}^2=\bar{m}_{\tilde{t}_R}^2=1000$\;GeV,
the left-right mixing parameter $X_t=A_t+\mu \cot\beta=500$\;GeV,
$\mu=200$\;GeV, $m_A=500$\;GeV and $\tan\beta=3$.
Masses are given in GeV.
\label{paraset}}
\begin{center}
\begin{tabular}{|c||c|c|c|c|c|c|} \hline
	Set&$\bar{m}_{\Phi_u^0}=\bar{m}_{\Phi_u^{\pm}}$& 
	$\bar{m}_{\Phi_d^0}=\bar{m}_{\Phi_d^{\pm}}$& 
	$\bar{m}_{\Omega^-}$&
	$\bar{m}_{\Omega^+}$&
	$\bar{m}_{\zeta}$&
	$\bar{m}_{\eta}$
	\\ \hline
	A&50&350&50&350&50&350
	\\
	B&50&400&50&400&50&400
	\\
	C&50&450&50&450&50&450
	\\
	\hline
\end{tabular}
\end{center}
\end{table}

The SM-like Higgs boson mass for benchmark parameter sets are shown in Figure 2.
Here we fix the parameters in the stop sector as $\bar{m}_{\tilde{t}_L}^2=\bar{m}_{\tilde{t}_R}^2=1000$\;GeV and 
the left-right mixing parameter as $X_t=A_t+\mu \cot\beta=500$\;GeV. 
The MSSM Higgs parameters are fixed as $\mu=200$\;GeV, $m_A=500$\;GeV and $\tan\beta=3$.
With this parameter set, the SM-like Higgs boson mass in the MSSM is evaluated as $m_h\simeq 102.3$\;GeV \cite{feynhiggs}.
In our model, the SM-like Higgs boson mass can get significant contributions 
 from loop diagrams involving the $Z_2$-odd fields due to the large coupling constant $\lambda$,
 in addition to the loop contributions from top quark fields.
The size of the corrections depends on the soft SUSY breaking parameters. 
As shown in Figure 2, the SM-like Higgs boson mass $m_h$ can reach to 125\;GeV by
the radiative corrections.

\begin{figure}[htbp]
\begin{center}
\includegraphics[width=90mm]{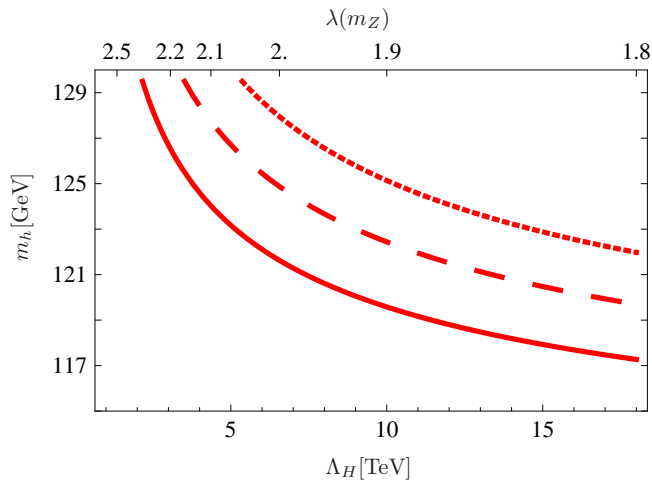}
\end{center}
\caption{The SM-like Higgs boson mass $m_h$. 
	The solid, dashed, and dotted curves correspond to the benchmark sets A, B, and C given in Table~\ref{paraset} 
	respectively.
	The MSSM parameters are fixed as  $\bar{m}_{\tilde{t}_L}^2=\bar{m}_{\tilde{t}_R}^2=1000$\;GeV, 
	$X_t=500$\;GeV, $\mu=200$\;GeV, $m_A=500$\;GeV and $\tan\beta=3$.
	}
	\label{figmh}
\end{figure}
In Table~\ref{lambench}, the values of $\Lambda_H$ and $\lambda$ corresponding to 
$m_h=125$\;GeV are displayed. Once these values are fixed, we can find out the 
mass spectrum for the $Z_2$-odd particles in each benchmark as shown in 
Table~\ref{mass}.
Since they are non-colored particles, linear colliders can have an advantage on 
the direct searches for them.

\begin{table}
\caption{The cutoff scale $\Lambda_H$ and the coupling constant $\lambda$ 
for realizing $m_h=125$\;GeV in each benchmark set.\label{lambench}}
\begin{center}
\begin{tabular}{|c||c|c|} \hline
	Set& $\Lambda_H$[TeV] &$\lambda (m_Z)$\\ \hline
	A&3.8&2.1
	\\
	B&6.4&2.0
	\\
	C&10.2&1.9
	\\
	\hline
\end{tabular}
\end{center}
\caption{The mass spectrum of the $Z_2$-odd particles for $m_h=125$\;GeV in 
each benchmark set.\label{mass} Masses are given in GeV.}
\begin{center}
\begin{tabular}{|c||c|c|c|c|c|c|c|c|c|c|c|c|} \hline
	Set&$m_{\Phi_u^{\pm}}$&$m_{\Phi_d^{\pm}}$&
	$m_{\Phi_u^{0}}$&$m_{\Phi_d^{0}}$&
	$m_{\Omega_-}$&$m_{\Omega_+}$&
	$m_{\zeta}$&$m_{\eta}$&
	$m_{\tilde{\chi}^{\prime \pm}_1}$&$m_{\tilde{\chi}^{\prime \pm}_2}$&
	$m_{\tilde{\chi}^{\prime 0}_1}$&$m_{\tilde{\chi}^{\prime 0}_2}$\\ \hline
	A&353.6&371.7&140.2&493.5&354.1&371.2&369.2&356.2&117.6&352.7&117.6&352.7
	\\
	B&331.9&417.2&134.2&516.0&332.5&416.7&414.9&334.7&110.3&331.0&110.3&331.0
	\\
	C&315.6&464.0&129.7&546.0&316.2&463.7&462.1&318.5&104.9&314.6&104.9&314.6
	\\ \hline
\end{tabular}
\end{center}
\end{table}

As shown in ref. \cite{4hdo}, 
 the F-terms from the couplings among the MSSM-like Higgs doublets, 
 the $Z_2$-odd doublets and the charged singlets can significantly enhance the first order electroweak phase transition
 if the coupling constant is as large as $\lambda \sim 2$.
The $Z_2$-odd neutral singlets can also contribute to making the phase transition stronger by the precise enhancement mechanism. 
Then the sphaleron decoupling condition required by successful electroweak baryogenesis is satisfied more easily.  
On the benchmark points chosen in the analysis, the first order electroweak phase transition is strong enough.
\\
\\

\section{Conclusions}

\ \ \ We have shown that the SUSY QCD theory with $N_c=2$ and $N_f=3$
 with one fundamental singlet $S$
 can give the strongly coupled 
 Higgs sector containing
 four iso-spin doublets, two charged singlets and two neutral singlets as the low-energy description.
Since the cutoff scale $\Lambda_H$ is as low as multi-TeV to 10 TeV,
 the coupling constant $\lambda$ in the Higgs sector is as large as $\sim 2$.
In our model, however,
 the SM-like Higgs boson is naturally light
 because the F-terms do not contribute to its mass at the tree level,
 while radiative corrections involving strongly coupled fields in the Higgs sector 
 are large enough to raise the SM-like Higgs boson mass to 125 GeV.

We comment on collider signatures of our model.
The model contains many new $Z_2$-odd charged and neutral scalars and fermions
 with masses of several hundred GeV.
The lightest one is definitely stable beacuse of the $Z_2$ parity.
\footnote{
The lightest $Z_2$-odd scalar and fermion can be both stable
 if their mass difference is smaller than the lightest $R$-parity odd, $Z_2$-even particle.
}
At the LHC, $Z_2$-odd particles are pair-produced through electroweak interactions,
 and decay into two lightest $Z_2$-odd particles, two lightest $R$-parity odd particles 
 and several SM particles. (For early studies, see refs.~\cite{early studies}.)
The most clear signatures of the model 
 are events with two or three leptons and large missing transverse momentum.
The results from the searches for slepton, chargino and neutralino direct productions
 by ATLAS collaboration \cite{collider} apply to our model.
The current bound \cite{collider} is mild and our benchmark spectra of Table 5
 may not have been excluded yet.

At the tree-level, 
 the SM-like Higgs boson couples to MSSM particles in the same way as the MSSM.
It also couples to $Z_2$-odd particles through large coupling constant $\lambda$,
 but its decay width and branching ratios at the tree-level are not altered
 if the mass of the light $Z_2$-odd particle is larger than half the SM-like Higgs boson mass.
The branching ratio into two photons
 can be significantly affected by loop corrections involving $Z_2$-odd charged scalars and fermions
 that strongly couples to the SM-like Higgs boson.
They can enhance or suppress the branching ratio depending on the parameters.
Also, the triple coupling for the SM-like Higgs boson, which has been studied in refs. \cite{hhh1, hhh2},
 receives large corrections from loops involving $Z_2$-odd doublets and singlets.
It thus significantly deviates from the SM prediction, and such deviation can be observed through future collider experiments.

Finally we comment on a possible extension of the model.
In the present model, 
 the large coupling constant $\lambda$ with the 125 GeV SM-like Higgs boson
 can make the first order electroweak phase transition strong enough to 
 enable electroweak baryongenesis \cite{4hdo}.
In addition, the lightest $Z_2$-odd field can be another source of dark matter
 than the lightest $R$-parity-odd field
 as long as it is electrically neutral.
Furthermore, 
 the model can be extended to explain the tiny neutrino masses.
By introducing $Z_2$-odd right-handed neutrino superfields
 whose Majorana masses are at the TeV scale,
 the tiny neutrino masses can be generated at loop levels \cite{ma, aks}.
With this extension of the model, we may be able to build the testable theory 
 in which baryon asymmetry of the Universe, dark matter and 
 the tiny neutrino masses can be simultaneously explained
 from the UV-complete SUSY strong dynamics around multi-TeV to 10 TeV
 without excessive fine-tuning.
We leave these topics for future studies.
\\
\\

\section*{Acknowledgements}

\ \ \ This work was supported in part by Grant-in-Aid for Scientific Research,
Nos. 22244031 (S.K.), 23104006 (S.K.), 23104011 (T.S.) and 24340046 (S.K. and T.S.).
The work of T.Y. was supported in part by a grant of the Japan Society for the Promotion of Science,
 No. 23-3599.


\begin{thebibliography}{99}
 
 
 \bibitem{lhc}
 ATLAS Collaboration, G. Aad et al., Phys. Lett. B710, 49 (2012) [arXiv:1202.1408[hep-ex]];
 CMS Collaboration, S. Chatrchyan et al. [arXiv:1202.1488 [hep-ex]].
 
 
 \bibitem{little higgs}
 N. Arkani-Hamed, A. G. Cohen and H. Georgi, Phys. Lett. B513, 232 (2001) 
 [hep-ph/0105239]; for a review, see
 M. Perelstein, Prog. Part. Nucl. Phys. 58, 247 (2007).
 
 
 \bibitem{mssm higgs}
 Y. Okada, M. Yamaguchi and T. Yanagida, Prog. Theor. Phys. 85, 1 (1991), Phys. Lett. B 262, 54 (1991);
 J. Ellis, G. Ridolfi and F. Zwirner, Phys. Lett. B 257, 83 (1991), Phys. Lett. B 262, 477 (1991);
 H. E. Haber and R. Hempfling, Phys. Rev. Lett. 66, 1815 (1991).
 
 
 \bibitem{hhh1}
 S. Kanemura, S. Kiyoura, Y. Okada, E. Senaha and C.-P. Yuan,
 Phys. Lett. B558, 157 (2003) [hep-ph/0211308];
 S. Kanemura, Y. Okada, E. Senaha and C.-P. Yuan,
 Phys. Rev. D70, 115002 (2004) [hep-ph/0408364].
 
 
 \bibitem{hhh2}
 S. Kanemura, T. Shindou and K. Yagyu,
 Phys. Lett. B 699, 258 (2011) [arXiv:1009.1836[hep-ph]].
 
 
 
 \bibitem{hhh-ewbg}
 C. Grojean, G. Servant and J. D. Wells,
 Phys. Rev. D71, 036001 (2005) [hep-ph/0407019];
 S. Kanemura, Y. Okada and E. Senaha,
 Phys. Lett. B606, 361 (2005) [hep-ph/0411354].
 
 
 \bibitem{4hdo}
 S. Kanemura, E. Senaha and T. Shindou,
 [arXiv:1109.5226[hep-ph]].
 
 
 
 \bibitem{sph dec}
 K. Funakubo and E. Senaha,
 Phys. Rev. D79, 115024 (2009) [arXiv:0905.2022[hep-ph]].
 
 
 \bibitem{ewbg}
 V. A. Kuzmin, V. A. Rubakov and M. E. Shaposhnikov, Phys. Lett. B155, 36 (1985); 
 A. G. Cohen, D. B. Kaplan and A. E. Nelson, Nucl. Phys. B349, 727 (1991);
 Ann. Rev. Nucl. Part. Sci. 43, 27 (1993); 
 M. Quiros, Helv. Phys. Acta 67, 451 (1994); 
 V. A. Rubakov and M. E. Shaposhnikov, Usp. Fiz. Nauk 166, 493 (1996) [Phys. Usp. 39 (1996) 461].
 
 
 \bibitem{ewbgmssm}
 M.~S.~Carena, M.~Quiros and C.~E.~M.~Wagner,
 Phys.\ Lett.\  B380, 81 (1996) [hep-ph/9603420];
 D.~Delepine, J.~M.~Gerard, R.~Gonzalez Felipe and J.~Weyers,
  Phys.\ Lett.\  B386, 183 (1996) [hep-ph/9604440];
 P.~Huet and A.~E.~Nelson,
  Phys.\ Rev.\  D53, 4578 (1996) [hep-ph/9506477];
 B.~de Carlos and J.~R.~Espinosa,
  Nucl.\ Phys.\  B503, 24 (1997) [hep-ph/9703212];
 M.~S.~Carena, M.~Quiros, A.~Riotto, I.~Vilja and C.~E.~M.~Wagner,
  Nucl.\ Phys.\  B503, 387 (1997) [hep-ph/9702409];
 M.~Aoki, A.~Sugamoto and N.~Oshimo,
  Prog.\ Theor.\ Phys.\ 98, 1325 (1997) [hep-ph/9706287];
 M.~Aoki, N.~Oshimo and A.~Sugamoto,
  Prog.\ Theor.\ Phys.\ 98, 1179 (1997) [hep-ph/9612225];
 J.~M.~Cline, M.~Joyce and K.~Kainulainen,
  JHEP0007, 018 (2000) [hep-ph/0006119];~
 M.~S.~Carena, J.~M.~Moreno, M.~Quiros, M.~Seco and C.~E.~M.~Wagner,
  Nucl.\ Phys.\  B599, 158 (2001) [hep-ph/0011055];
 M.~S.~Carena, M.~Quiros, M.~Seco and C.~E.~M.~Wagner,
  Nucl.\ Phys.\  B650, 24 (2003) [hep-ph/0208043];
 C.~Lee, V.~Cirigliano and M.~J.~Ramsey-Musolf,
  Phys.\ Rev.\  D71, 075010 (2005) hep-ph/0412354];
 V.~Cirigliano, M.~J.~Ramsey-Musolf, S.~Tulin and C.~Lee,
  Phys.\ Rev.\  D73, 115009 (2006) [hep-ph/0603058];
 T.~Konstandin, T.~Prokopec, M.~G.~Schmidt and M.~Seco,
  Nucl.\ Phys.\  B738, 1 (2006) [hep-ph/0505103];
 D.~J.~H.~Chung, B.~Garbrecht, M.~J.~Ramsey-Musolf and S.~Tulin,
  Phys.\ Rev.\ Lett.\ 102, 061301 (2009) [arXiv:0808.1144 [hep-ph]];
 K.~Funakubo, S.~Tao and F.~Toyoda,
  Prog.\ Theor.\ Phys.\ 109, 415 (2003) [hep-ph/0211238];
 M.~Carena, G.~Nardini, M.~Quiros and C.~E.~M.~Wagner,
  Nucl.\ Phys.\  B812, 243 (2009) [arXiv:0809.3760 [hep-ph]].
 
 

 
 \bibitem{fat}
 R. Harnik, G. D. Kribs, D. T. Larson and H. Murayama,
 Phys. Rev. D 70, 015002 (2004) [hep-ph/0311349].
  
  
 \bibitem{beyond fat1}
  S.~Chang, C.~Kilic and R.~Mahbubani,
  Phys.\ Rev.\ D71, 015003 (2005)  [hep-ph/0405267];
  A.~Delgado and T.~M.~P.~Tait,
  JHEP0507, 023 (2005)  [hep-ph/0504224].


 \bibitem{beyond fat2}
  C. Liu, Phys. Rev. D61, 115001 (2000) [hep-ph/9910303];
  M. A. Luty, J. Terning and A. K. Grant, Phys. Rev. D63, 075001 (2001) 
  [hep-ph/0006224];
  H. Murayama, [hep-ph/0307293].



 \bibitem{nMSSM}
 C. Panagiotakopoulos and K. Tamvakis, 
 Phys. Lett. B446, 224 (1999) [hep-ph/9809475];
 C. Panagiotakopoulos and K. Tamvakis, 
 Phys. Lett. B469, 145 (1999) [hep-ph/9908351];
 C. Panagiotakopoulos and A. Pilaftsis, Phys. Rev. D63, 055003 (2001) [hep-ph/0008268];
 A. Dedes, C. Hugonie, S. Moretti and K. Tamvakis, Phys. Rev. D63, 055009 (2001)
 [arXiv:hep-ph/0009125].


 \bibitem{Int-Sei}
 K. A. Intriligator and N. Seiberg, Nucl. Phys. Proc.
Suppl. 45BC, 1 (1996) [arXiv:hep-th/9509066].


 \bibitem{nda}
 H. Georgi, A. Manohar and G. W. Moore, Phys. Lett.
B 149, 234 (1984); 
 H. Georgi and L. Randall, Nucl.
Phys. B 276, 241 (1986); 
 M. A. Luty, Phys. Rev. D
57, 1531 (1998) [arXiv:hep-ph/9706235]; 
 A. G. Cohen, D. B. Kaplan and A. E. Nelson, Phys. Lett. B 412, 301
(1997) [arXiv:hep-ph/9706275].


 
 \bibitem{techful}
 H. Murayama,
 [arXiv:hep-ph/0307293].
 
 
 \bibitem{NMSSM}
 For a review, see

U. Ellwanger, C. Hugonie and A. M. Teixeira,
Phys. Rept. 496, 1 (2010) [arXiv:0910.1785 [hep-ph]].


 \bibitem{feynhiggs}
 T. Hahn, S. Heinemeyer, W. Hollik, H. Rzehak and G. Weiglein,
 Nucl. Phys. Proc. Suppl. 205-206: 152-157 (2010) [arXiv:1007.0956 [hep-ph]];
 
 
 \bibitem{early studies}
 R. Barbieri, L. J. Hall and V. S. Rychkov,
 Phys. Rev. D 74, 015007 (2006)
 [arXiv:hep-ph/0603188];
 Q. H. Cao, E. Ma and G. Rajasekaran,
 Phys. Rev. D 76, 095011 (2007)
 [arXiv:0708.2939 [hep-ph]].
 
 
 \bibitem{collider}
 ATLAS Collaboration,
 [arXiv:1208.2884 [hep-ex]];
 ATLAS Collaboration,
 [arXiv:1208.3144 [hep-ex]];


 \bibitem{ma}
 E. Ma, Phys. Rev. D 73, 077301 (2006) [hep-ph/0601225];
 J. Kubo, E. Ma and D. Suematsu, Phys. Lett. B 642, 18 (2006) [hep-ph/0604114];
 E. Ma, Annales Fond. Broglie 31, 285 (2006) [hep-ph/0607142];
 H. Fukuoka, J. Kubo and D. Suematsu, Phys. Lett. B 678, 401 (2009) [arXiv:0905.2847 [hep-ph]];
 E. Ma, Mod. Phys. Lett. A23, 721 (2008) [arXiv:0801.2545 [hep-ph]]; 
 D. Suematsu and T. Toma, Nucl. Phys. B847, 567 (2011) [arXiv:1011.2839 [hep-ph]].


 \bibitem{aks}
 M.~Aoki, S.~Kanemura and O.~Seto,
  Phys.\ Rev.\ Lett.\ 102, 051805 (2009)  [arXiv:0807.0361 [hep-ph]];
 M.~Aoki, S.~Kanemura and O.~Seto,
  Phys.\ Rev.\ D80, 033007 (2009)  [arXiv:0904.3829 [hep-ph]];
 M.~Aoki, S.~Kanemura and K.~Yagyu,
  Phys.\ Rev.\ D83, 075016 (2011)  [arXiv:1102.3412 [hep-ph]].
 
 
\end{thebibliography}
\end{document}